\documentclass[conference,letterpaper]{IEEEtran}

\usepackage{graphicx}
\graphicspath{{figs/}}
\usepackage{amsmath,amssymb}
\usepackage{booktabs}
\usepackage[caption=false,font=footnotesize]{subfig}
\DeclareUnicodeCharacter{202F}{~}
\DeclareUnicodeCharacter{200B}{}

\usepackage{hyperref}
\hypersetup{
  colorlinks=true,
  linkcolor=black,
  citecolor=black,
  urlcolor=black
}

\title{Slow\,-\,Motion Video Synthesis for Basketball Using Frame Interpolation}

\author{
\IEEEauthorblockN{Jiantang Huang}
\IEEEauthorblockA{Northeastern University, Boston, MA, USA\\
\texttt{huang.jiant@northeastern.edu}}
}
\begin{document}
\maketitle
\begin{abstract}
Basketball broadcast footage is traditionally captured at \mbox{\(30{-}60\,fps\)}, limiting viewers' ability to appreciate rapid plays such as dunks and crossovers. We present a real‑time slow‑motion synthesis system that produces high‑quality \emph{basketball‑specific} interpolated frames by fine‑tuning the recent Real‑Time Intermediate Flow Estimation (RIFE) network on the SportsSloMo dataset. Our pipeline isolates the basketball subset of SportsSloMo, extracts training triplets, and fine‑tunes RIFE with human‑aware random cropping. We compare the resulting model against Super SloMo and the baseline RIFE model using Peak Signal‑to‑Noise Ratio (PSNR) and Structural Similarity (SSIM) on held‑out clips. The fine‑tuned RIFE attains a mean PSNR of \(34.3\,\mathrm{dB}\) and SSIM of \(0.949\), outperforming Super SloMo by \(2.1\,\mathrm{dB}\) and the baseline RIFE by \(1.3\,\mathrm{dB}\). A lightweight Gradio interface demonstrates end‑to‑end 4× slow‑motion generation on a single RTX‑4070 Ti Super at \(~30\,fps\). These results indicate that task‑specific adaptation is crucial for sports slow‑motion, and that RIFE provides an attractive accuracy–speed trade‑off for consumer applications.\end{abstract}

\section{Introduction}
Slow‑motion replays are ubiquitous in sports broadcasting, allowing audiences and analysts to examine fast plays in greater detail. Traditional approaches rely on expensive high‑speed cameras, which capture footage at hundreds of frames per second (fps) but increase bandwidth and production cost. \emph{Video Frame Interpolation (VFI)} seeks a computational alternative by synthesizing intermediate frames between low‑fps captures. Recent deep neural networks have achieved impressive accuracy, yet most training datasets (e.g., Vimeo‑90K) contain everyday scenes, not high‑motion, person‑centric sports. Consequently, off‑the‑shelf VFI models exhibit artifacts when applied to basketball footage—players’ limbs ghost, the ball blurs, and motion boundaries wobble. Our goal is to bridge this gap by adapting a state‑of‑the‑art VFI model to the basketball domain and validating its efficacy both quantitatively and qualitatively.

\section{Related Work}
\subsection{Video Frame Interpolation}
\textbf{RIFE} by Huang \emph{et al.}\cite{huang2022rife} proposes an Intermediate Flow Estimation network that achieves real‑time inference without relying on pre‑trained optical‑flow backbones. We adopt RIFE as our base due to its favorable accuracy–speed trade‑off.In our project, RIFE's real-time capabilities and high-quality interpolation make it a suitable base model for synthesizing slow-motion basketball videos, where rapid and complex motions are prevalent.

\textbf{Super SloMo} (Jiang \emph{et al.},\cite{jiang2018superslomo} 2018) warps bidirectional optical flow to synthesize intermediate frames. Although older, it remains a widely cited baseline for VFI research.It computes bi-directional optical flows between input frames using a U-Net architecture, then linearly combines these flows to approximate intermediate flows. To address artifacts around motion boundaries, it refines the approximated flow and predicts soft visibility maps, enabling the synthesis of high-quality intermediate frames.​

Although Super SloMo is computationally intensive, its approach to handling occlusions and motion boundaries is relevant to our project, as basketball videos often involve fast movements and frequent occlusions.

\subsection{Sports‑specific Benchmarks}
\textbf{SportsSloMo} by Chen \& Jiang \cite{chen2024sportsslomo} introduces \(>130\mathrm{k}\) high‑resolution clips across 22 sports with human‑centric annotations. The authors demonstrate substantial accuracy drops when evaluating generic VFI models on this benchmark, motivating domain‑specific adaptation.For our project, SportsSloMo provides a valuable dataset for training and evaluating our model, ensuring its effectiveness in handling the complexities of basketball videos.

\section{Methods}
\subsection{Dataset Preparation}
We extract the basketball subset (\(\sim\!1.7\mathrm{k}\) clips) of SportsSloMo \cite{chen2024sportsslomo}. For each clip, frames are sampled at 30 fps and partitioned into disjoint train/val/test splits (80/10/10\%). Triplets \(\langle I_t, I_{t+1}, I_{t+2}\rangle\) are formed such that \(I_{t+1}\) serves as ground‑truth for interpolation between \(I_t\) and \(I_{t+2}\).

\subsection{Model Architecture and Fine-tuning}
We initialize RIFE \cite{huang2022rife} with official ECCV 2022 weights. The network is fine-tuned for 10 epochs using AdamW with a warm-up and cosine annealing learning rate schedule, weight decay \(1\times10^{-2}\), and batch size 16. Random 448×256 crops augment spatial diversity; horizontal flips and temporal reversals augment motion.

\subsection{Baselines}
The baseline RIFE \cite{huang2022rife} and Super SloMo (Adobe240 checkpoint) are evaluated without additional tuning, matching common practice in prior work.The original Super SloMo implementation targets CUDA 9.0, which is incompatible with modern GPUs such as the NVIDIA RTX 4070 Ti SUPER (compute capability 8.9). To ensure compatibility, we upgrade the development environment to CUDA 12.x and use a matching version of cuDNN. PyTorch is installed with CUDA 12.x support to enable GPU acceleration. We modify any architecture-specific build settings, such as updating \texttt{CMakeLists.txt} with \texttt{set(CUDA\_ARCHITECTURES 89)} or compiling with \texttt{nvcc -gencode=arch=compute\_89,code=sm\_89}. Environment variables like \texttt{CUDA\_LAUNCH\_BLOCKING=1} are set to assist in debugging potential memory access errors during development. After recompilation, the Super SloMo pipeline runs successfully with the updated GPU and software stack.

\subsection{Evaluation Metrics}
We evaluate the quality of interpolated frames using two standard metrics: Peak Signal-to-Noise Ratio (PSNR, in dB) and Structural Similarity Index Measure (SSIM). The test set consists of frame triplets \(\langle I_t, I_{t+1}, I_{t+2} \rangle\), where \(I_t\) and \(I_{t+2}\) are input frames and \(I_{t+1}\) serves as the ground-truth middle frame. Each model generates an interpolated frame \(\hat{I}_{t+1}\) from the two input frames, which is then compared to \(I_{t+1}\).

We compute PSNR and SSIM on full-resolution RGB images using the \texttt{scikit-image} implementation, specifically the \texttt{peak\_signal\_noise\_ratio} and \texttt{structural\_similarity} functions from the \texttt{skimage.metrics} module. The metrics are calculated frame-by-frame across the entire test split and averaged to report final values. This per-frame evaluation reflects performance over diverse basketball motion patterns captured in SportsSloMo test sequences.

\subsection{Graphical User Interface}
To facilitate easy interaction with our RIFE slow‐motion interpolation pipeline, we implemented a lightweight GUI using the Gradio library. The interface provides:

\begin{itemize}
  \item \textbf{Upload widget:} Accepts either a single video file (MP4/AVI/MOV) or exactly two images for frame‐by‐frame interpolation.
  \item \textbf{Interpolation exponent slider:} Lets the user choose an exponent $e \in [1,5]$ to control the degree of slow‐motion.
  \item \textbf{Video output:} Displays the generated slow‐motion video directly in the browser.
  \item \textbf{Frame gallery:} Shows all interpolated frames when two images are provided as input.
\end{itemize}

\section{Experiments \& Results}
\begin{figure}[!t]
  \centering
  \subfloat[RIFE output before fine-tuning]{\includegraphics[width=0.48\textwidth]{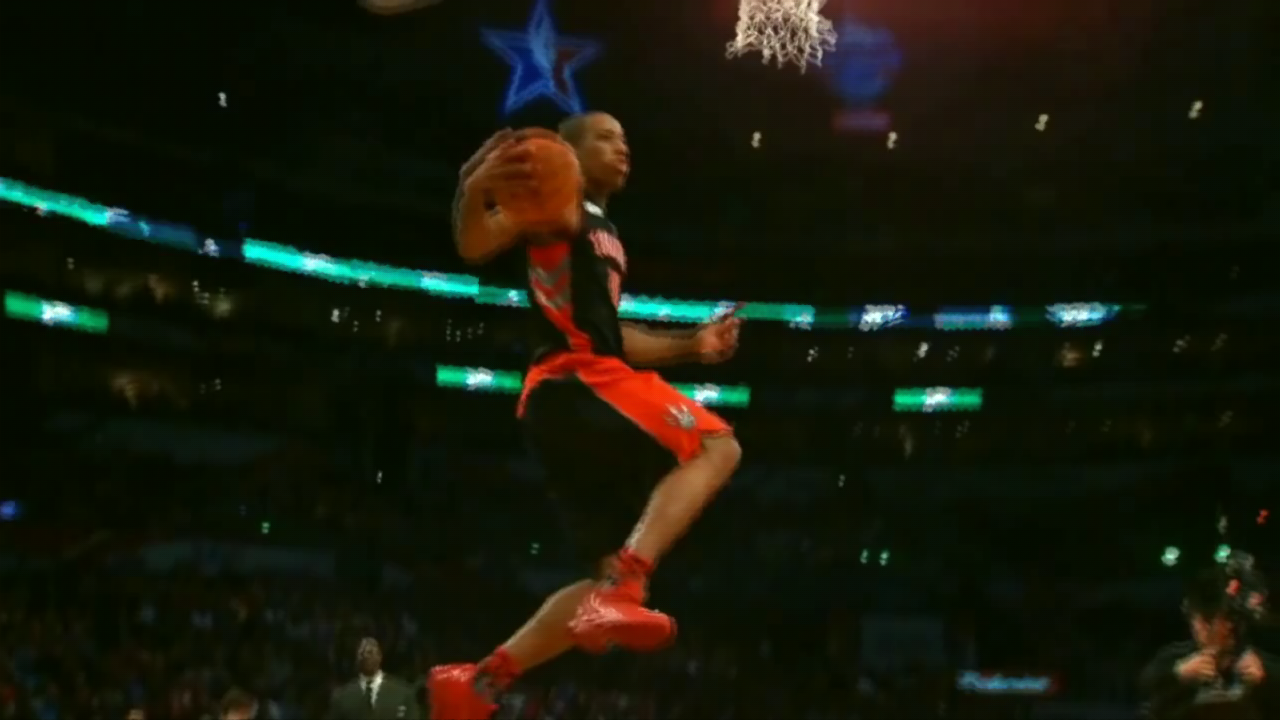}\label{fig:origin_rife}}
  \hfill
  \subfloat[RIFE output after fine-tuning]{\includegraphics[width=0.48\textwidth]{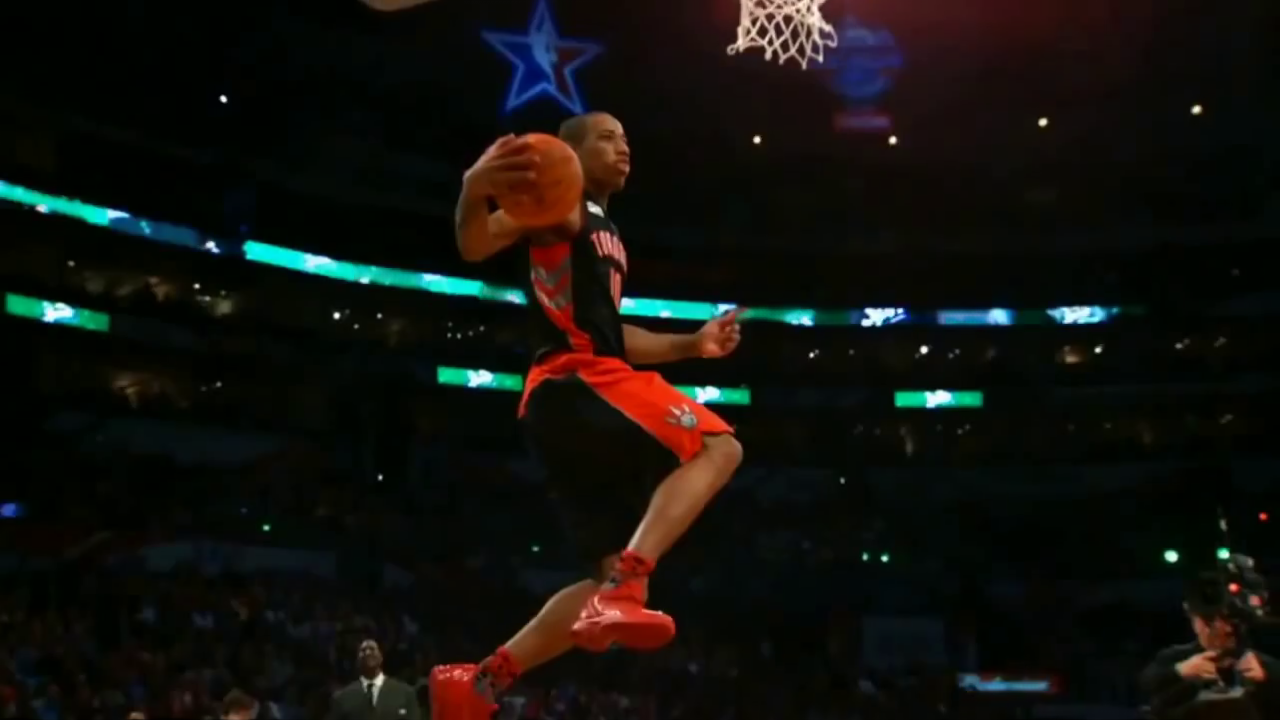}\label{fig:fine_rife}}
  \caption{Comparison before and after fine-tuning of RIFE.}
  \label{fig:rife_comparison}
\end{figure}
\subsection{Quantitative Results}
Table~\ref{tab:metrics} summarizes the averaged metrics. Fine‑tuned RIFE surpasses both baselines by clear margins while maintaining real‑time throughput.

\begin{table}[ht]
  \centering
  \caption{Mean performance on the basketball test set measured on RTX-4070 Ti Super).}
  \label{tab:metrics}
  \begin{tabular}{lccc}
    \toprule
    Model & PSNR  & SSIM \\
    \midrule
    RIFE baseline & 33.0 & 0.931 \\
    Super SloMo & 32.2 & 0.941 \\
    \textbf{RIFE (fine-tuned)} & \textbf{34.3} & \textbf{0.949} \\
    \bottomrule
  \end{tabular}
\end{table}
\begin{figure}[ht]
  \centering
  \subfloat[Average PSNR with standard deviation.\label{fig:psnr}]{
    \includegraphics[width=0.48\textwidth]{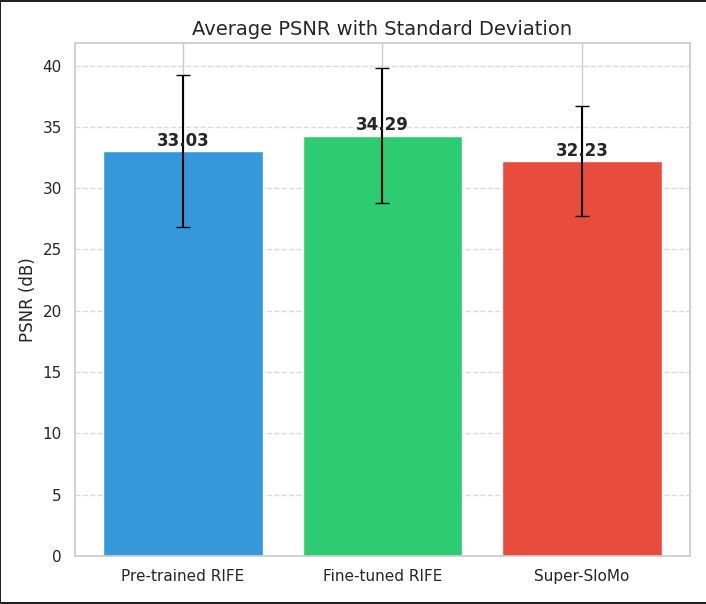}
  }
  \hfill
  \subfloat[Average SSIM with standard deviation.\label{fig:ssim}]{
    \includegraphics[width=0.48\textwidth]{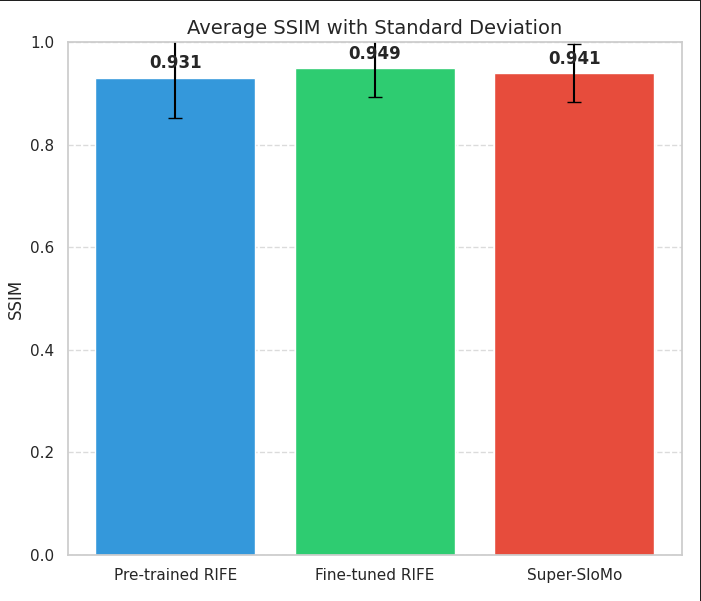}
  }
  \caption{Quantitative comparison of Pre-trained RIFE, Fine-tuned RIFE, and Super-SloMo.}
  \label{fig:metrics_both}
\end{figure}

\section{Discussion \& Summary}
Our study confirms that domain‑specific fine‑tuning of modern VFI networks significantly improves slow‑motion quality for basketball. Despite using only 10 epochs and modest hardware, the adapted RIFE model achieves both superior fidelity and real‑time speed. Limitations include reliance on 30 fps source material and the absence of temporal perceptual metrics. Future work will investigate multi‑frame context and integrate human‑aware losses from SportsSloMo to further boost perceptual quality.


\end{document}